\documentclass{aa}  
\usepackage{graphicx,natbib,psfig}  
\newcommand{\ltsima} {$\; \buildrel < \over \sim \;$}  
\newcommand{\gtsima} {$\; \buildrel > \over \sim \;$}  
\newcommand{\lta} {\lower.5ex\hbox{\ltsima}}  
\newcommand{\gta} {\lower.5ex\hbox{\gtsima}}

\newcommand{\kms}{$\rm{\,km \,s}^{-1}$}
\begin{document}  

\title{The supermassive black hole in the Seyfert 2 galaxy NGC 5252 \thanks
{Based  on observations obtained at
the  Space  Telescope Science  Institute,  which  is  operated by  the
Association of  Universities for Research  in Astronomy, Incorporated,
under NASA contract NAS 5-26555.}} 
  
\titlerunning{The supermassive black hole in NGC 5252}  
  
\author{Alessandro Capetti
\inst{1}
\and    
Alessandro Marconi \inst{2}    
\and   
Duccio Macchetto \inst{3,4}
\and   
David Axon \inst{5}}   
   
\offprints{A. Capetti}  
     
\institute{INAF - Osservatorio Astronomico di Torino, Strada
  Osservatorio 20, I-10025 Pino Torinese, Italy\\
\email{capetti@to.astro.it}
\and 
INAF - Osservatorio Astrofisico di Arcetri
       Largo E. Fermi 5, I-50125 Firenze, Italy\\
\email{marconi@arcetri.astro.it}
\and
Space Telescope Science Institute
3700 San Martin Drive, Baltimore, MD 21218, USA
\and
Affiliated with  ESA's Space Telescope Division\\
\email{macchetto@stsci.edu}
\and
Department of Physics,
Rochester Institute of Technology, 85 Lomb Memorial Drive,
Rochester, NY 14623, USA\\
\email{djasps@rit.edu}}

\date{}  
   
\abstract{We present  results from HST/STIS  long-slit spectroscopy of
the gas  motions in  the nuclear  region of the  Seyfert 2  galaxy NGC
5252. The observed  velocity field is  consistent with gas  in regular
rotation  with  superposed localized  patches  of  disturbed gas.  The
dynamics  of the  circumnuclear gas  can be  accurately  reproduced by
adding to the stellar mass component  a compact dark mass of $M_{BH} =
0.95  ~   (-0.45;+1.45)  \times   10^9$  M$_{\sun}$,  very   likely  a
supermassive  black  hole  (BH).  Contrarily to  results  obtained  in
similar studies rotational broadening  is sufficient to reproduce also
the behaviour of  line widths.  The BH mass estimated  for NGC 5252 is
in  good agreement  with the  correlation between  $M_{BH}$  and bulge
mass. The  comparison with the $M_{BH}$ vs  $\sigma_c$ relationship is
less  stringent   (mostly  due  to  the  relatively   large  error  in
$\sigma_c$); NGC  5252 is located above  the best fit  line by between
0.3 and 1.2  dex, i.e. 1 - 4 times the  dispersion of the correlation.
Both the  galaxy's and  BH mass of  NGC 5252 are  substantially larger
than those  usually estimated for  Seyfert galaxies but, on  the other
hand,  they  are  typical   of  radio-quiet  quasars.   Combining  the
determined BH  mass with the  hard X-ray luminosity, we  estimate that
NGC 5252 is emitting at a  fraction $\sim$ 0.005 of L$_{Edd}$. In this
sense, this active nucleus appears  to be a quasar relic, now probably
accreting at a low rate, rather than a low black hole mass counterpart
of a  QSO.   } \maketitle
  
\section{Introduction}
\label{intro}

It is widely accepted that Active Galactic Nuclei (AGN) are powered by
accretion of matter onto massive  black holes.  AGN activity peaked at
$z\sim1-2$ and  the high ($>10^{12}$L$_{\sun}$)  luminosities of quasi
stellar objects  (QSOs) are explained by accretion  onto super massive
($\sim  10^8-10^{10}$  M$_{\sun}$) black  holes  at  or  close to  the
Eddington limit.  The  observed evolution of the space  density of AGN
(e.g. So\l tan 1982, Marconi et al. 2004, Yu \& Tremaine 2002) implies
that  a significant  fraction  of luminous  galaxies  must host  black
holes, relics of past activity.  Indeed,  it is now clear that a large
fraction of hot  spheroids contain a massive BH  (Kormendy \& Gebhardt
2001, Merritt  \& Ferrarese 2001) and  it appears that the  BH mass is
proportional  to  both  the   mass/luminosity  of  the  host  spheroid
(Magorrian  et  al. 1998,  Marconi  \&  Hunt  2003) and  its  velocity
dispersion (Ferrarese \& Merritt  2000, Gebhardt et al. 2000, Tremaine
et al. 2002).

Several radio-galaxies, all associated with giant elliptical galaxies,
like M87 (Macchetto et al.  1997),  M84 (Bower et al.  1998), NGC 7052
(van der Marel \& van den  Bosch 1998) and Centaurus A (Marconi et al.
2001),  are  now  known  to  host supermassive  ($\sim  10^8  -  10^9$
M$_{\sun}$)  BHs in  their  nuclei. The  luminosity  of their  optical
nuclei  indicates that they  are accreting  at a  low rate  and/or low
accretion efficiency  (Chiaberge, Capetti  \& Celotti 1999),  with the
exception  of  Cygnus  A  (Tadhunter  et al.  2003).  They  presumably
sustained quasar  activity in  the past but  at the present  epoch are
emitting much below their Eddington  limits (L / L$_{Edd} \sim 10^{-5}
- 10^{-7}$.)

For the radio-quiet AGN, there are two viable explanations for the low
luminosity  of Seyfert galaxies  compared to  QSO.  Firstly,  they may
represent a population of less massive BH's accreting at a level close
to their  Eddington limit. Support for this  interpretation comes from
the  estimated BH mass  derived from  reverberation mapping  which are
found to be in the  range between $\simeq 10^6$ M$_{\sun}$ and $\simeq
10^8$ M$_{\sun}$ (Wandel, Peterson \&  Malkan 1999, Kaspi et al. 2000,
Peterson 2003).   However the  data are far  from extensive and  it is
still possible that in some Seyfert nuclei there are much more massive
BH, currently in  a dormant phase in which the  accretion rate is much
reduced from their hay-day as QSO's, i.e.  they are relics of past QSO
activity of the sort described  above. In practice a hybrid population
is likely  to exist  in galaxies as  a whole  as a consequence  of the
regulation of BH  growth by fluctuations in accretion  rate with time.
The  study  of  the  Seyfert  BH mass  distribution  then  provides  a
statistical  method of investigating  the interplay  between accretion
rate  and BH  growth.  In  order to  achieve this  it is  necessary to
directly  measure (bound)  the BH  masses in  Seyfert galaxies  and to
compare  their Eddington  and Bolometric  luminosities using  the hard
X-ray luminosities.

Clearly, measurements  of BH masses in Seyfert  galaxies would provide
us not  only with very important insights  concerning their connection
with the past quasars activity,  but represent also a crucial test for
their unified  scheme, as e.g.  systematic differences  between the BH
measured in the two spectroscopic types will argue against the general
validity  of  the unified  model.   Similarly  important  will be  the
comparison between the BH masses  found in Seyfert galaxies with those
of  non active galaxies.   Several ongoing  HST/STIS program  aimed at
measuring BH masses  in spiral and disk galaxies  should constrain the
mass function and space density of BHs, and their connection with host
galaxy  properties (e.g., Hubble  type, bulge  and disk  mass, central
velocity dispersion, etc).   A comparison of the BH  masses in Seyfert
and  those  derived  from  these  programs will  allow  us  to  define
differences between  quiescent and active galaxies.   In particular it
will  be possible to  test whether  the current  level of  activity is
accompanied by significant growth of the BH mass in active galaxies.

To date,  the handful  of measurements of  BH masses in  Seyferts have
been  obtained using essentially  the reverberation  mapping technique
(Wandel, Peterson  \& Malkan  1999, Kaspi et  al. 2000,  Wandel 2002).
Despite lingering worries about the limitation of this technique (e.g.
Krolik  2001)  the good  agreement  between  these  estimates and  the
predictions of  the $\sigma_c$ -  M$_{BH}$ and $M_{bulge}$  - M$_{BH}$
correlations supports the idea that they are indeed reliable estimates
of M$_{BH}$  (e.g. Wandel 2002 ,  Peterson 2003).  However,  this method is
limited to  Seyfert 1 only  as the broad  line region in Seyfert  2 is
hidden from our  view.  Estimates of M$_{BH}$ using  H$_2$O masers have not
provided  similarly compelling  evidence or  accurate  measurements in
Seyfert 2 galaxies (Greenhill et  al.  1996, 1997a, 1997b, 2003) as in
the classic case of NGC4258 (Miyoshi et al. 1995).

Leaving  aside  the reverberation  mapping  technique,  to detect  and
measure the masses of massive BHs requires spectral information at the
highest possible  angular resolution --  the "sphere of  influence" of
massive BHs is typically \ltsima 1\arcsec\ in radius even in the closest
galaxies.  
Nuclear absorption line  spectra can be used to demonstrate
the presence  of a BH, but  the interpretation of the  data is complex
because it involves stellar-dynamical models that have many degrees of
freedom  (e.g. Valluri, Merritt, \& Emsellem 2003).  
In  Seyfert galaxies  the  problems are
compounded  by  the  copious  light  from the  AGN.   Studies  at  HST
resolution  of  ordinary optical  emission  lines  from  gas disks  in
principle provide a more widely applicable and readily interpreted way
of detecting BHs (cf. Macchetto et al.  1997, Barth et al.  2001)
provided  that  the  gas  velocity  field are  not  dominated  by  non
gravitational motions. 
When dealing with luminous AGN, this is a particularly important
issue, as the active nucleus might substantially
affect the nuclear gas dynamics.
In general,  black hole mass estimates based on
gas dynamics  have large uncertainties  if the inclination of  the gas
disk cannot be  constrained at the low inclination  end. A significant
advantage of  studying Seyfert 2  galaxies is that the  unified scheme
requires  the innermost  regions to  be seen  at a  sufficiently large
inclination to  allow the circumnuclear  torus to obscure  its nucleus
and Broad Line Region, effectively eliminating this problem.

Following this  chain of reasoning,  we obtained STIS  observations of
NGC 5252, an early type (S0) Seyfert  2 galaxy at a redshift z = 0.023
in order to study its nuclear  kinematics and to determine the mass of
its central black hole. Adopting 
H$_{\rm o}$ = 75 Km s$^{-1}$ Mpc$^{-1}$, 
at  the  distance  of NGC  5252  (92 Mpc),  1\arcsec
corresponds to 450 pc.

Line emission  in this galaxy shows a  remarkable biconical morphology
(Tadhunter \& Tsvetanov 1989) extending out to 20 kpc from the
nucleus along  PA -15$^\circ$. 
On  a sub-arcsec scale  (Tsvetanov et  al.  1996),
three emission line knots form a linear structure oriented at PA $\sim
35$, close  to the bulge major  axis. 
Ground  based measurements of  the large
scale velocity field  of the gas were obtained by  Morse et al. (1998)
with a  resolution of 1.4$\arcsec$. They found that, while at radii
larger than 40$\arcsec$ the gas rotates in the plane of the stellar
disk, at smaller radii there is evidence for two superposed 
dynamical components: 
a gaseous disk significantly inclined (by $\sim 40^\circ$)
with respect to the galaxy's plane and traced by a the dusty spiral
structure identified by Tsvetanov et al. (1996); a counterrotating
ring coplanar with the stellar disk.

The paper  is organized as follows:  in Sec. \ref{obs}  we present the
observations and the data reduction that lead to the results described
in Sec.  \ref{results}.  In Sec.  \ref{fitting} we model  the observed
rotation curves  taking into account  the star distribution.   We will
show  that the  dynamics of  the circumnuclear  gas can  be accurately
reproduced  by circular  motions  in  a thin  disk.   Our results  are
discussed in Sec. \ref{discussion}.  

\section{Observations and data reduction}
\label{obs}

\begin{figure}
\centering
\psfig{figure=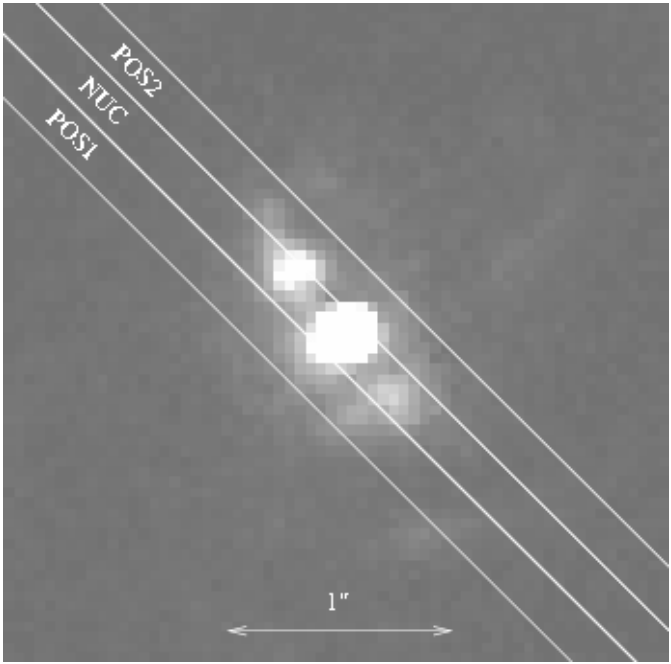,width=1.\linewidth,angle=-90}
\vskip 0.5cm
\caption{\label{slits} H$\alpha$ image of the central 3\arcsec $\times$
3\arcsec of NGC 5252 with superposed the three slit locations.
North is up, East is left.} 
\end{figure}

STIS observations were obtained on Jan 29th, 1999.  The nucleus of NGC
5252 was  acquired with the ACQ  mode with two 60  s exposures through
the F28X50LP optical long-pass  filter.  The 50\arcsec x 0\farcs2 slit
was then  positioned at 3  different locations: on the  continuum peak
(position NUC) and at two off-nuclear positions (OFF1 and OFF2) offset
by $\pm$ 0\farcs2.  The orientation  of the slit was -135$^\circ$ from
North, approximately  aligned with the galaxy's major axis
and the triple inner  structure of line
emission.  Fig.  \ref{slits} shows  the slit locations superposed onto
the WFPC2 H$\alpha$ image.

At each position, we used the medium dispersion grating G750M,
which provides a scale of 0.56 \AA ~ pixel$^{-1}$
and, in combination with the 0\farcs2 slit, a resolution of 
$\mathcal{R} = 3000$, 
centered on the redshifted H$\alpha$ line and covering the rest frame
wavelength range 6160-6700 \AA. 
The pixel size in the spatial direction is 0\farcs0507.
The exposure time was set at 1000 s for OFF1 and OFF2 and
at 1400 s on the nucleus.

The data were calibrated using the CALSTIS pipeline to perform the steps 
of bias subtraction, dark subtraction, applying
the flat field, and combining the two subexposures to reject 
cosmic-ray events. To reject hot pixels from the data, 
we employed dark frames obtained immediately before our
observations. The data were then wavelength-calibrated and rectified 
by tracing the wavelenght calibration lamps 
and then applying these solutions for the
geometric distortions to the data. 

We then selected the spectral regions containing the lines of interest and
subtracted the continuum by a polynomial/spline fit pixel by pixel along the
dispersion direction. The continuum subtracted lines were fitted, row by row,
along the dispersion direction with Gaussian functions using the
task SPECFIT in STSDAS/IRAF. All emission lines present in the spectra
(H$\alpha$, [N II]$\lambda\lambda$6548,6584 
and [O I]$\lambda\lambda$6300,6363) were fitted simultaneously with the same
velocity and width.  

\begin{figure*}
\centerline{
\psfig{figure=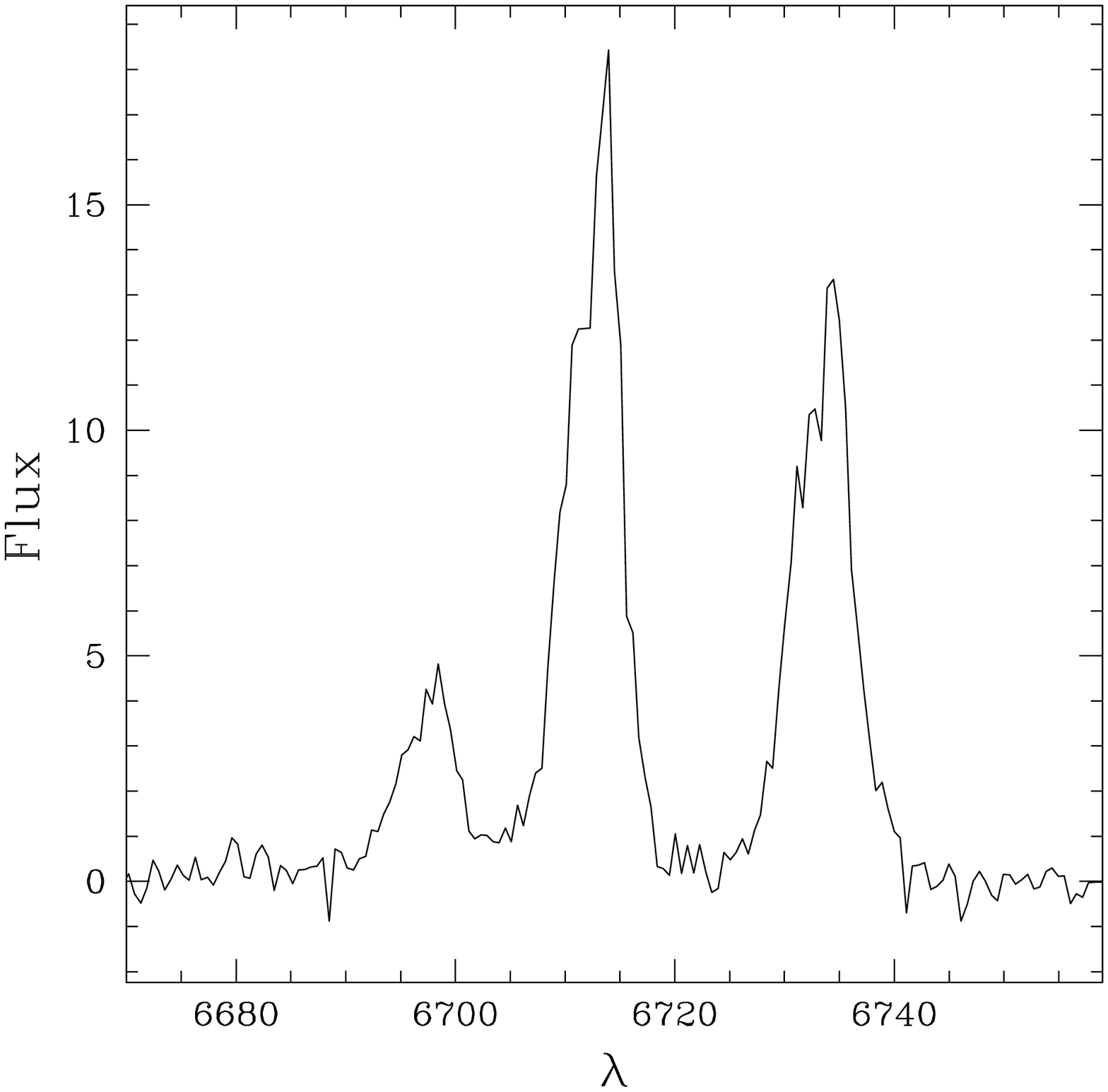,width=0.5\linewidth}
\psfig{figure=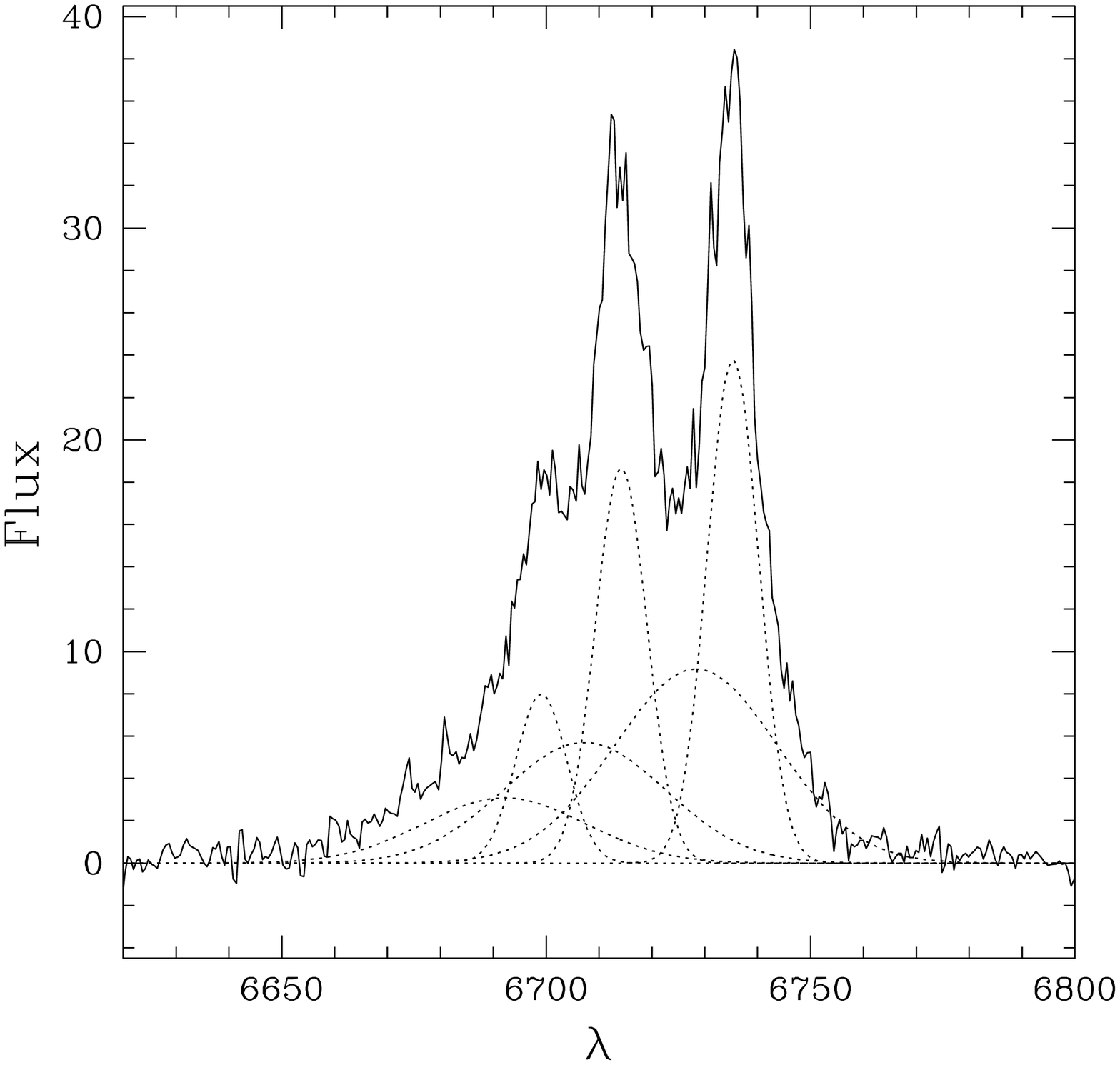,width=0.5\linewidth}}
\vskip 0.5cm
\caption{\label{profiles} Emission lines profiles
on the nucleus (right panel) and at 0\farcs3 NE of the nucleus
(left panel) showing the regions were the line profile cannot be fitted with
a single Gaussian and a high velocity component was added.}
\end{figure*}

In three regions a single Gaussian does not produce 
an accurate fit to the line profile: on the nucleus,
where there is a strong blue wing, 
and in the range 0\farcs2 $<$ r $<$ 0\farcs4 on both sides of the nucleus
in the NUC slit (at the inner edges of the off-nuclear knots)  
where the line profile shows a broad base
offset from the narrow line core (see Fig. \ref{profiles}). 
Two Gaussians fit were performed in these regions.
We preferred to model the blue wing in the nuclear 
spectrum with a high velocity component
rather than with a broad $H\alpha$ line as this line asymmetry
is also seen in the [O I] and [S II] lines. This contrasts 
the classification of NGC 5252 as a Sy 1.9 by Osterbrock \& Martel (1993).

Where the
SNR was insufficient the fitting was improved by co-adding two or
more pixels along the slit direction.  

\section{Results}
\label{results}

\begin{figure}
\centering
\psfig{figure=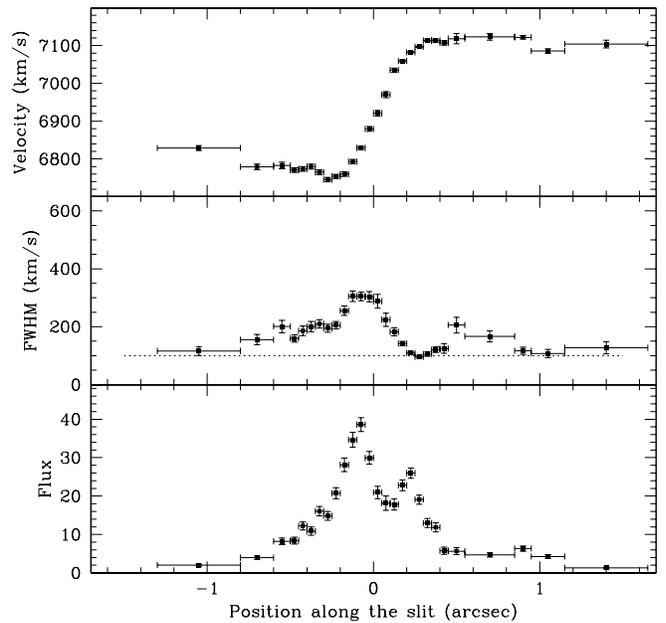,width=1.0\linewidth}
\caption{\label{pos1} Velocity, FWHM and flux at along the POS1 slit. 
Positions along the slit are relative to the continuum peak, 
positive values are SW. The instrumental line-width of 100 \kms\ is drawn  
as a dotted line in the FWHM panel.}
\vskip 0.5cm
\end{figure}

\begin{figure}
\centering
\psfig{figure=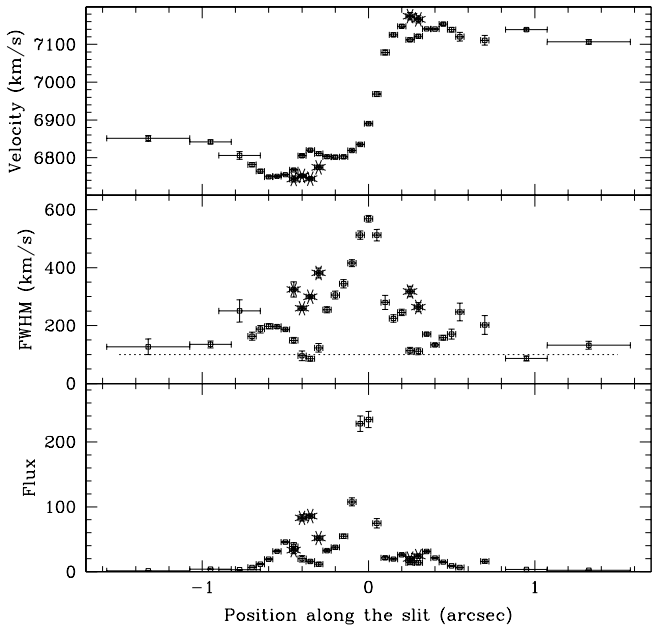,width=1.0\linewidth}
\caption{\label{nuc} Same as Fig. \ref{pos1} for the NUC slit. 
The patches of disturbed
gas, spatially coincident with the off-nuclear line knots, are 
marked with asterisks. Here lines are fitted with a double Gaussian
to separate the broad perturbed component from the rotating gas.}
\vskip 0.5cm
\end{figure}

\begin{figure}
\centering
\psfig{figure=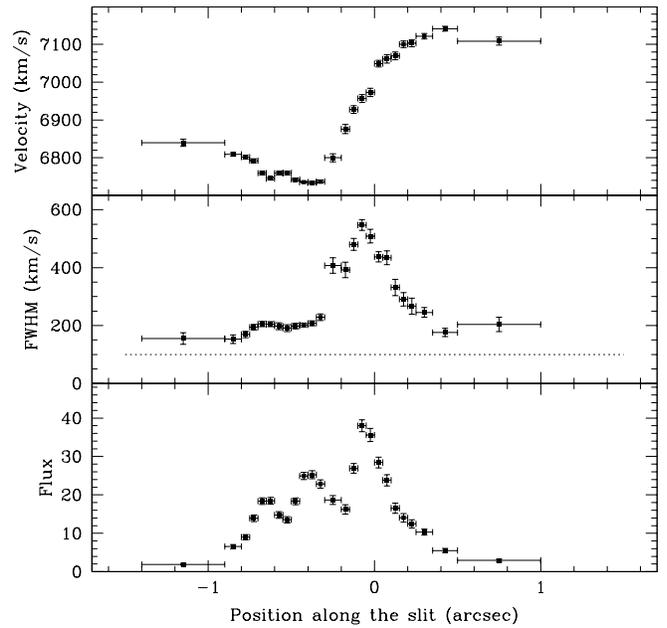,width=1.0\linewidth}
\caption{\label{pos2} Same as Fig. \ref{pos1} for the POS2 slit.}
\vskip 0.5cm
\end{figure}

The results obtained from the fitting procedure 
at the three slit positions are shown in Fig. \ref{pos1} through
Fig. \ref{pos2}
where we show the line central velocity, flux and FWHM at each location
along the slits. Emission is detected out to a radius of
$\sim$ 1\farcs6 corresponding to $\sim$ 720 pc.

On the central slit, the line emission (Fig. \ref{nuc}, central panel) 
is strongly concentrated showing
a bright compact knot cospatial with the continuum peak.
Two secondary emission line maxima are also present 
at $\pm$ 0.35\arcsec\ from the main peak.
They represent the intersection of the slit with emission line blobs seen in 
Fig. ~ \ref{slits}. 

The velocity curve has a full amplitude of
$\sim$ 400 \kms and it shows a general reflection symmetry.
Starting from the center of symmetry (at $r$ $\sim$ 0\farcs1
and $v$ $\sim$ 6950 \kms) 
the velocity rapidly rises on both sides by $\sim$ 200 \kms reaching a
peak a $r$ $\sim$ 0\farcs2. In this central region both the line flux and
the line width rapidly
decrease (from the maximum value of 600 \kms\ to 100 \kms).

In correspondence with the off-nuclear knots ($r  \sim$ 0\farcs3 -
0\farcs6)  the situation is more complex
as in these regions two velocity components are present.
The narrower component (FWHM $\sim 100-150$ \kms) 
has a lower (and decreasing) velocity offset.
Its width and flux also decrease with radius, 
following the trend seen at smaller radii.
The second component, marked with an asterisk in Fig. \ref{nuc},
is considerably broader (FWHM $\sim 250-400$
\kms) and shows higher velocity
offset with respect to the center of symmetry. At radii larger than
$r$ $\sim$ 0\farcs4 the intensity of the narrow component 
falls below our detection
threshold and only the broader component is visible. Lines remain
broad out to  $\sim$ 0\farcs8 from the nucleus. At even larger distances
the velocity field flattens.

It therefore appears that two different gas systems are present
in the nuclear regions of NGC 5252: 
the first shows a symmetric velocity field, with decreasing line width 
and can be interpreted as being produced by gas in rotation around the 
nucleus, counter-rotating with respect to the large scale stellar and
gas disk. The second component, showing significant 
non circular motions, is found to be  
associated exclusively with the off-nuclear blobs.

The velocity fields in the off-nuclear positions are quite similar to the one
seen on the nucleus, but here the signal-to-noise 
is insufficient to provide detailed information on the line
profiles. The presence of the high-velocity gas 
is nonetheless clearly revealed
looking at the behaviour of the line width 
in correspondence of the line knots, but the broad and narrow
components cannot be separated.

\section{Modeling the rotation curves}
\label{fitting}

Our modeling code, described in detail in Marconi et al. (2003),
was used to fit the observed rotation curves. Very briefly the code computes
the rotation curves of the gas assuming that the gas is rotating
in circular orbits within a thin disk in the galaxy potential. 
The gravitational potential has two components: the stellar potential 
(determined in the next section), characterized by its mass-to-light ratio 
and a dark mass concentration (the black hole), 
spatially unresolved at HST+STIS resolution and
characterized by its total mass M$_{BH}$.  
In computing the rotation curves we take
into account the finite spatial resolution of HST+STIS,
the line surface brightness and we integrate over
the slit and pixel area. The $\chi^2$ is minimized to determine the 
free parameters using the downhill simplex algorithm by Press et al. (1992).

\subsection{The stellar mass distribution}
\label{stelle}

In order to account for the contribution of the stars to the gravitational
potential in the nuclear region, the stellar luminosity density is
derived from the observed surface brightness distribution.  

\begin{figure}
\centering
\psfig{figure=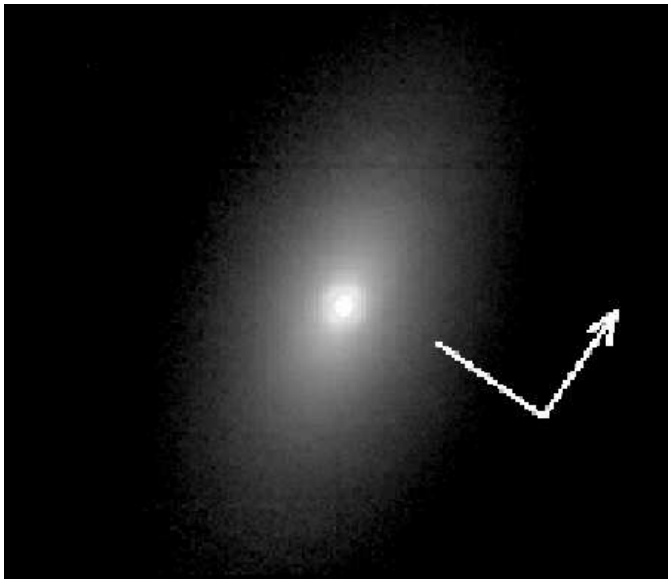,width=1.\linewidth,angle=0}
\caption{\label{nicmos} Central 8 $\times$ 8 arcsec of the NICMOS F160W 
image of NGC 5252. The compass indicates the North-East orientation.
Note the smooth regular shape of the galaxy and the bright central 
unresolved source.}
\end{figure}

\begin{figure}
\centering
\psfig{figure=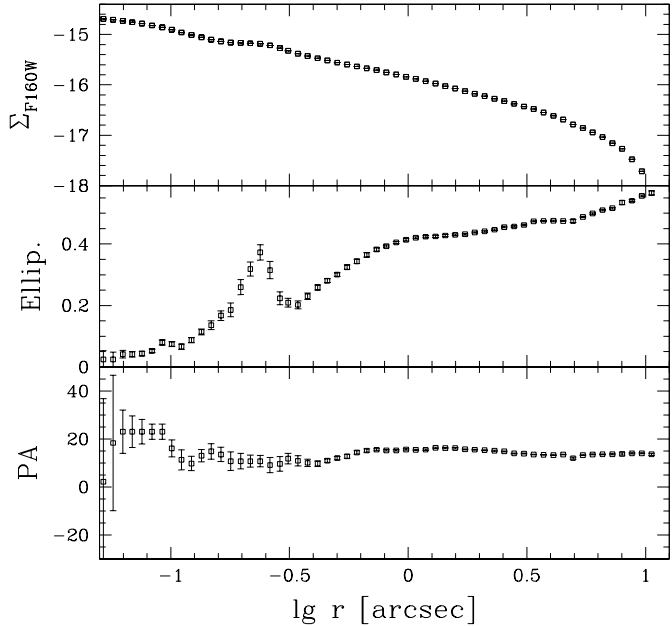,width=1.\linewidth,angle=0}
\vskip 0.5cm
\caption{\label{ellipse} Results of the isophotes analysis of the H band
images of NGC 5252. Surface brightness is shown in the top panel
(in units of erg s$^{-1}$ cm$^{-2}$ \AA$^{-1}$ arcsec$^{-2}$), the 
galaxy's ellipticity and position angle are shown in the middle and 
bottom panels respectively.}
\end{figure}

We reconstructed the galaxy light profile using two NICMOS
F160W (H band) images obtained with the 
NIC1 and NIC2 cameras whose pixel size are 
0\farcs043 and 0\farcs075 respectively.
The first image, with smaller pixel size 
(shown in Fig. \ref{nicmos}), was used for the
central regions (r $<$ 1\arcsec) while for the more extended emission
we took advantage of the larger field of view of camera NIC2.
We used the IRAF/STSDAS program ELLIPSE to fit elliptical isophotes
to the galaxy that is particularly well behaved
(see Fig. \ref{ellipse}).
Excluding the nuclear regions that are dominated by a central 
unresolved source, the ellipticity is essentially constant at a value of 0.45
out to a radius of 6\arcsec where it slowly increases to a value of 0.55.
Also the position angle does not show significant variations
being approximately constant at PA = 15$^\circ$. 
All isophotes are concentric within half a NIC1 pixel, i.e. 0\farcs02.

\begin{figure}
\centering
\psfig{figure=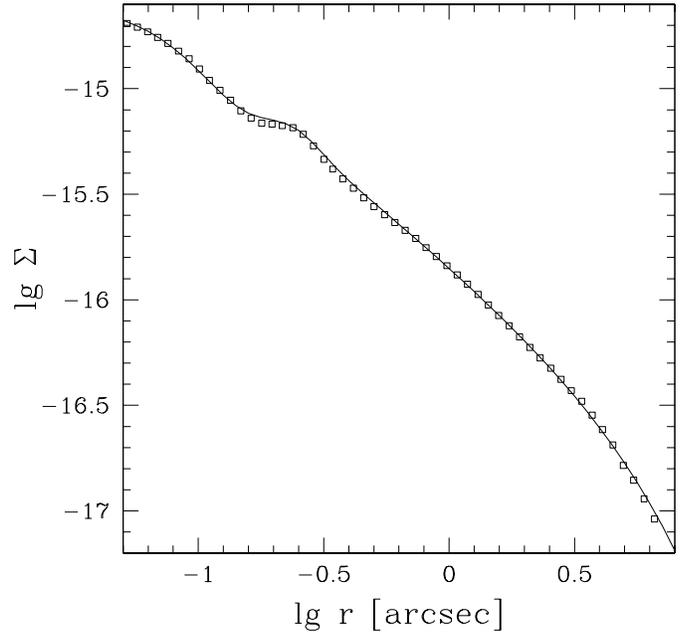,width=1.\linewidth,angle=0}
\vskip 0.5cm
\caption{\label{mass} Brightness profile of NGC 5252. Surface brightness is
in units of erg s$^{-1}$ cm$^{-2}$ \AA$^{-1}$ arcsec$^{-2}$. Fit to the 
profile obtained from an oblate spheroid stellar density distribution 
with an added nuclear point source.}
\end{figure}

The inversion procedure to derive the stars distribution
from the surface brightness 
is not unique if the gravitational potential does not have a
spherical symmetry as in the case in NGC 5252 where the 
isophotes are not circular.  
Assuming that the gravitational potential is an
oblate spheroid, the inversion depends on the knowledge of the potential
axial ratio $q$, and the inclination of its principal plane with respect 
to the line of sight. As these two quantities are related by the observed 
isophote ellipticity, we are left with the freedom of assuming different 
galaxy inclinations to the line of sight.  
The measured ellipticity implies a minimum inclination
for the galaxy of 53$^\circ$. We therefore performed the deprojection
for three representative values of the galaxy inclination, namely 55$^\circ$, 
70$^\circ$ and 85$^\circ$. 
Following van der Marel \& van den Bosch (1998), we 
assumed an oblate spheroid density distribution parameterized as:
$$
\rho(m) = \rho_0\left(\frac{m}{r_b}\right)^{-\alpha} \left[1+\left(\frac{m}{r_b}\right)^2\right]^{-\beta}
$$
$m$ is given by
$m^2 = x^2+y^2+z^2/q^2$
where $xyz$ is a reference system with the $xy$ plane corresponding to the
principal plane of the potential and $q$ is the intrinsic axial ratio.  
A detailed description of the relevant
formulas and of the inversion and fit procedure is presented in the
Marconi et al. (2003). 
The best fit obtained for an inclination of 70$^\circ$ 
is shown in Fig. \ref{mass} 
with $\alpha$ = 2.0, $\beta$=3.1 and $r_b$ =17.5\arcsec.
The presence of an unresolved  nuclear point source, and its
associated Airy ring, are clearly visible
in Fig. \ref{nicmos}. Indeed a point source  with flux (7.3 $\pm$
0.1) ~10$^{-17}$ erg s$^{-1}$ cm$^{-2}$ \AA$^{-1}$ had to be added to
the  extended luminosity  distribution to  provide a  good fit  to the
brightness profile. It  has been shown by Quillen  et al.  (2001) that
unresolved infrared  sources are  found in the  great majority  of HST
images  of  Seyfert  galaxies   and  that  their  luminosities  strongly
correlate with  both the hard X-ray  and the [O  III] line luminosity.
This result suggests  a dominant AGN contribution to  the IR emission.
This seems to be the case also for the nuclear source of  NGC 5252 (that
is part of the Quillen et al. sample) as its
luminosities in  the IR  (1.2$\times 10^{42}$ erg  s$^{-1}$) and  hard X-ray
($\sim  2 \times 10^{43}$  erg s$^{-1}$,  Cappi et  al. 1996)  make its
representative point to  lie on the IR -  X-ray correlation defined by
Seyfert nuclei. 

\subsection{Fitting the gas kinematics}

Our modeling code was used to fit the nuclear rotation curve. 
Clearly data points associated to the
high velocity components in the region 0\farcs2 $<r<$ 0\farcs8 
on both sides of the nucleus, must be excluded from the fitting
procedure, but there is some level of arbitrariness in the decision of
exactly which portion of the rotation curve to use for the dynamical 
modeling. As we showed in Sec. \ref{results}, based on the behaviour
of gas velocity, line width and flux, it appears that it is possible
to isolate the component extending the regular nuclear velocity field
from the perturbed high velocity patches, at least within
0\farcs2 $<r<$ 0\farcs4 from the nucleus. 
Nonetheless, we prefer to always examine the robustness of our results 
against the more conservative choice of excluding altogether this critical
spatial region.
For the off-nuclear slits the high velocity component associated
  to the off-nuclear knots is clearly also
present (see e.g. the local velocity minima at r=-0\farcs3 and -0\farcs6,
corresponding to maxima of flux in Fig. \ref{pos2}) but here the
signal to noise is insufficient to deblend it from the narrow component.
We conservatively used only the 5 five central
points of OFF1 and OFF2 and those at radii larger than 0\farcs8. 
The fit was also constrained using the 
values of the line width over the same regions.

The observed emission line surface brightness was modeled with a composition
of two Gaussians, the first reproducing the central emission peak
while the second accounts for the brightness behaviour at large radii.
This brightness distribution was used to build the synthetic kinematic models.

Having fixed the line brightness distribution, 
the free parameters of the fit are 
\begin{itemize}
\item
the systemic velocity, $v_{sys}$, 
\item
the impact parameter (i.e. the distance between
the nuclear slit center and the center of rotation) $b$,
\item
the position of the galaxy center along the nuclear slit $s_0$,
\item
the angle between the slits and the line of nodes, $\theta$, 
\item
the disk inclination $i$,
\item
the mass to light ratio, $\Upsilon$, 
\item
the black hole mass $M_{BH}$.

\end{itemize}

Concerning the disk inclination, in a oblate spheroid, 
the stable orbits of the gas are coplanar
with the principal plane of the potential and it is possible 
to directly associate the galaxy inclination and line of nodes with 
those of the circumnuclear gas.
However, the potential shape is not sufficiently 
well determined by the isophotal fitting
down to the innermost regions of the galaxy and it is quite possible that
a change of principal plane might occur at the smallest radii, in particular
in the presence of a supermassive black hole. We then preferred to 
leave the disk inclination as a free parameter of the fit.

We then performed a $\chi^2$ minimization allowing all 
parameters to vary freely. 
We also tested that the effects of assuming different galaxy inclinations 
for the deprojection of the stars surface brightness have a negligible
impact on the final values of the free parameters.

\begin{figure*}
\centerline{
\psfig{figure=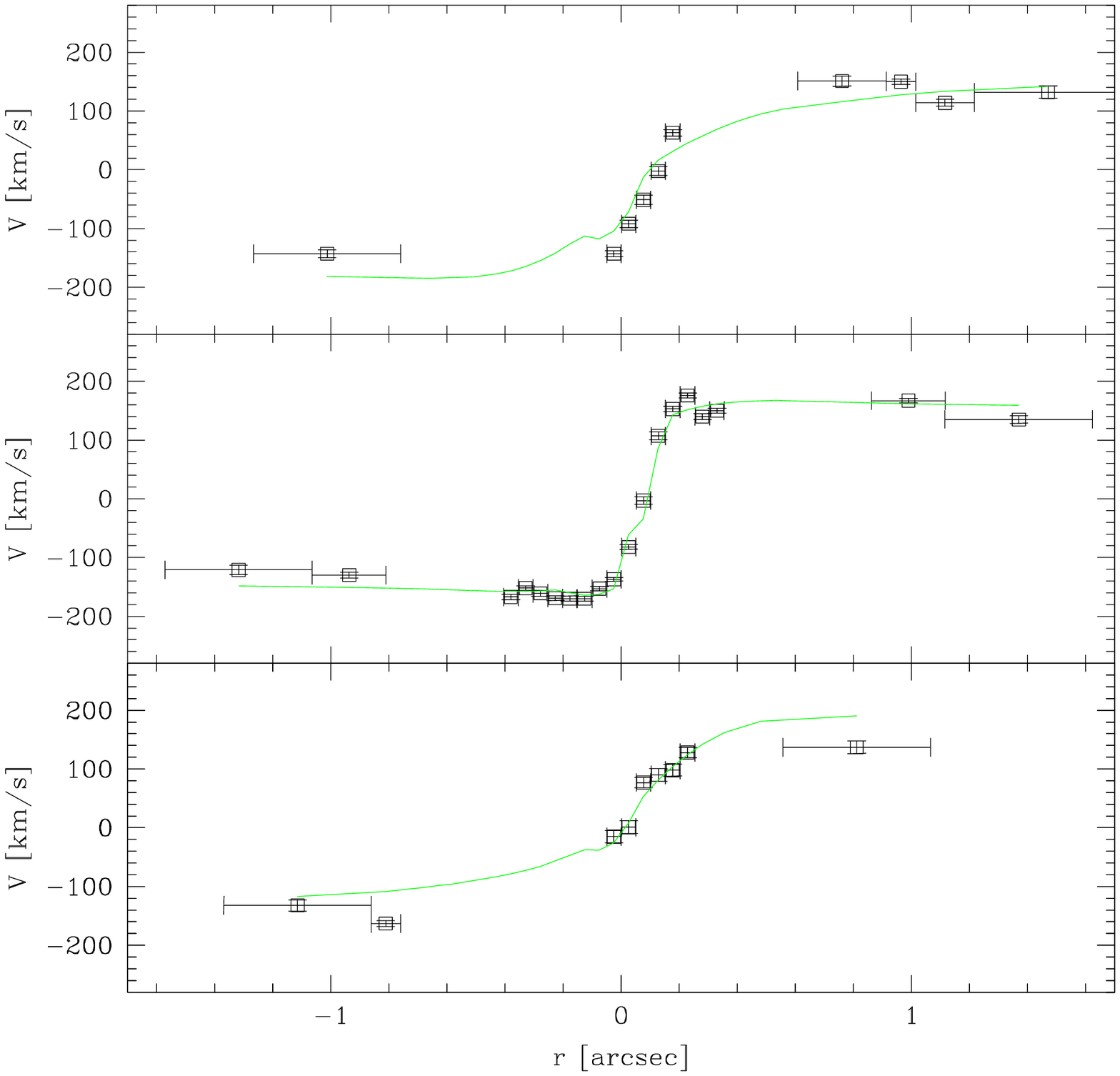,width=0.5\linewidth,angle=0}
\psfig{figure=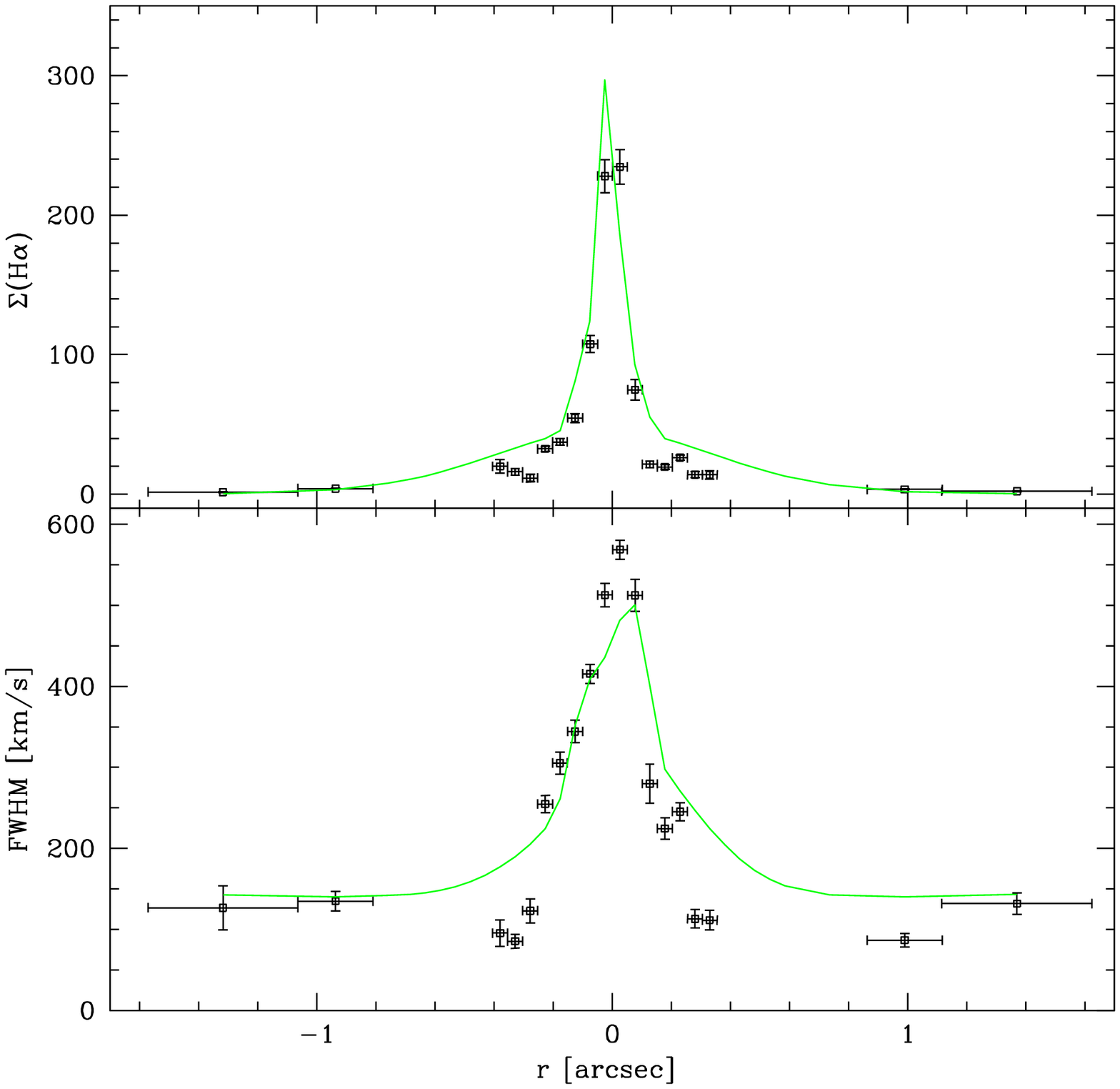,width=0.5\linewidth,angle=0}}
\vskip 0.5cm
\caption{\label{fit70} Best fit to the rotation curves (left),
line surface brightness (upper right panel) 
and line width (lower right panel).}
\end{figure*}

The best fit to our data is shown in Fig.
\ref{fit70} and is obtained for the set of parameters reported in Table 1.
The residuals between the fit and the data (see Fig. \ref{res}, upper panel)
show the good agreement between the model and the observed velocity points
and line width 
indicating that the kinematics of gas in the circumnuclear
regions of NGC 5252 can be successfully accounted for by circular
motions in a thin disk. We also note that, contrarily to 
results obtained in similar studies 
(Barth et al. 2001, Verdoes Kleijn et al. 2000, 2002), 
no additional source of internal gas speed   
is needed, in addition to rotational or instrumental broadening,
to reproduce the behaviour of line width.

\begin{table*}
\caption{Best fit parameter set}
\begin{tabular}{l c c c c c c c} \hline
$i$   & $b$ & $s_0$ & $\theta$ & $V_{sys}$ & $\Upsilon$ & M (M$_{\sun}$) & $\chi^2_r$ \\ 
\hline
77 &  -0.03 & 0.01 & 193 & 6950 &  0.46 & 0.95$\times10^9$ & 16.5 \\
\hline
\end{tabular}
\label{bestfit}
\end{table*}

To estimate the uncertainty associated to the black hole mass 
estimate we explored its variation with respect to the other free
parameters, and in particular with those that are more strongly
coupled to it, the gas inclination $i$ and the mass-to-light ratio
$\Upsilon$.

For the gas inclination, 
in the case of NGC 5252 the orientation of the galaxy is
constrained to be larger than 53$^\circ$. On the other hand,
even if a change of orientation occurs at small radii, 
since NGC 5252 is classified as a Seyfert 2, 
invoking the unified scheme imposes, as we described in the Introduction,
a tight lower bound on the inclination of $i > 30$ (very conservative).
We then selected two representative
values, 45$^\circ$ and 70$^\circ$, for our analysis. 

At these fixed values of $i$ we derived 
the best fit for each pair of BH mass and $\Upsilon$ values 
leaving all other parameters free to vary. 
We then build a $\chi^2$ grid at varying $\Upsilon$ and BH mass. 

The values of minimum
$\chi^2$ are far larger than the values indicative of
a good fit and this is in contrast with the fact that the curves shown in the
figures seem to trace the data points well.  The reason for this discrepancy is
that $\chi^2$ is not properly normalized (e.g. 
because not all points are independent or as they do not
include the uncertainties in the relative wavelength calibration 
for the three slits) and/or imply the presence
of small deviations from pure rotation.
Following Barth et al. (2001) we then rescaled the error bars by 
adding in quadrature a constant
error such that best fitting model provides $\chi^2$/dof $\sim$ 1.
This is a quite conservative approach as it has the effect of 
increasing the final uncertainty on $M_{BH}$.
The additional error are found to be 23 km s$^{-1}$ for the velocity and
80 km s$^{-1}$ for the velocity width.
We rescaled all values of $\chi^2$ with this procedure.

\begin{figure}
\centerline{
\psfig{figure=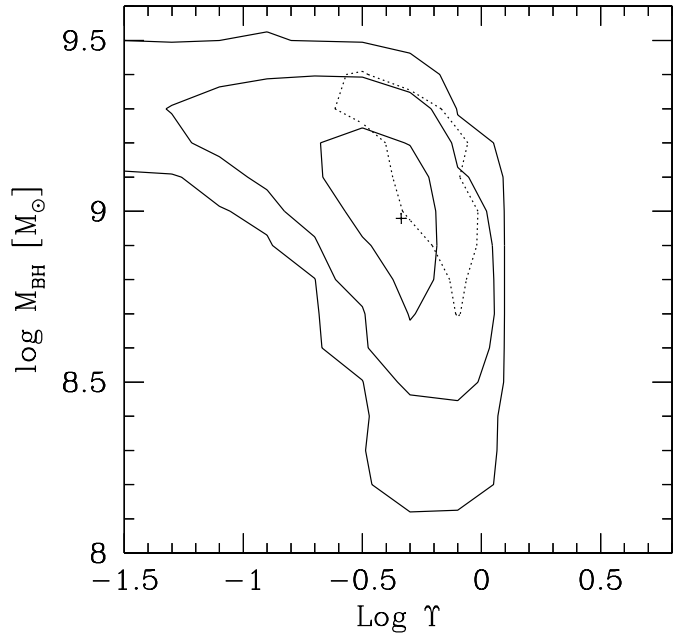,width=1.0\linewidth,angle=0}}
\vskip 0.5cm
\caption{\label{rescale} $\chi^2$ contours at varying
$\Upsilon$ and  M$_{BH}$ 
for an inclination of 70$^\circ$.
Contours are plotted for confidence levels of 1, 2 and 3 $\sigma$.
Plotted in a dotted contour is the 1 $\sigma$ confidence level
for 45$^\circ$, while the plus sign mark the overall best fit.}
\end{figure}

The result of this analysis is presented in Fig. \ref{rescale}
where we plot the confidence levels. 
The overall 1 $\sigma$ range of the BH mass is 
$M_{BH} = 0.95_{-0.45}^{+1.45}\times 10^9$ M$_{\sun}$.

At this stage we tested  how the exclusion of the velocity points
  in  the  region  0\farcs2   $<r<$  0\farcs4  affects  this  estimate
  re-running  the  modeling  code. The
  difference from the  above estimate is marginal as  we found a BH
  mass only $\sim$ 10 \% larger ($1.07 \times 10^9$ M$_{\sun}$) with an 
uncertainty increased by $\sim$ 15 \%.

We also estimated the allowed range for the remaining free parameters
looking at their range of values restricting to the region of the 
$\Upsilon$ vs BH mass plane within the 1 $\sigma$ confidence level.
All parameters are remarkably stable.
\begin{itemize}

\item
The mass to light ratio in the H band $\Upsilon$ ranges from  
0.27 to 0.93, values characteristic of 
a stellar population with an age of a few Gyr (see Maraston 1998).
Only the lower end of this range corresponds to ages
which are uncomfortably low for an S0 galaxy; however, given the relatively
weak correlation between $M_{BH}$ and $\Upsilon$, at least within the
1 $\sigma$ confidence region, excluding a posteriori low values of 
mass to light ratios would not significantly affect our
estimates of the black hole mass.

\item
The angle between the slit and the disk line of nodes is 
15$^\circ$ $\pm$ 10$^\circ$, corresponding to a position angle of the 
disk major axis of 30$^\circ\pm$ 10$^\circ$ to be compared with host
galaxy's major axis at 15$^\circ$. This is 
consistent with the idea that,
although counterrotating with respect to the stars,
the nuclear gas is coplanar with the
galaxy disk as expected for settled orbits in the galaxy's potential,
As already mentioned in the Introduction, ground based studies 
evidenced the presence of two gaseous structure in the central regions
of NGC 5252. The velocity field seen by HST appears to be
extension at the smallest scale of the counterrotating ring.

\item
The systemic velocity found from our fit is 
6950 $\pm$ 10 km s$^{-1}$ (with an additional uncertainty of 
12  km s$^{-1}$ associated to the absolute wavelength calibration)
and it is consistent within the errors with the value 
of 6916 $\pm$ 42 NGC 5252 reported by Falco et al. (1999).

\end{itemize}

\section{Discussion}
\label{discussion}

Our model fitting of the nuclear rotation curve of NGC 5252 
indicates that the kinematics of gas in its innermost regions
can be successfully accounted for by circular motions in a thin disk
when a point-like dark mass (presumably a supermassive black hole)
of $M_{BH} = 0.95\times10^9$ M$_{\sun}$ is added to the 
galaxy potential.

Let us firstly explore how this mass determination 
is connected with the properties of the host galaxy.

Concerning the connection between black hole and bulge mass, 
Marconi \& Hunt (2003) report a bulge mass for NGC 5252 of 2.4$\times10^{11}$
M$\sun$. The value expected for M$_{BH}$ in NGC 5252 from their 
correlation between
M$_{BH}$ and M$_{bul}$ is 5.7 $~10^8$ M$_{\sun}$ in close agreement
with our estimate. 

\begin{figure*}
\centerline{
\psfig{figure=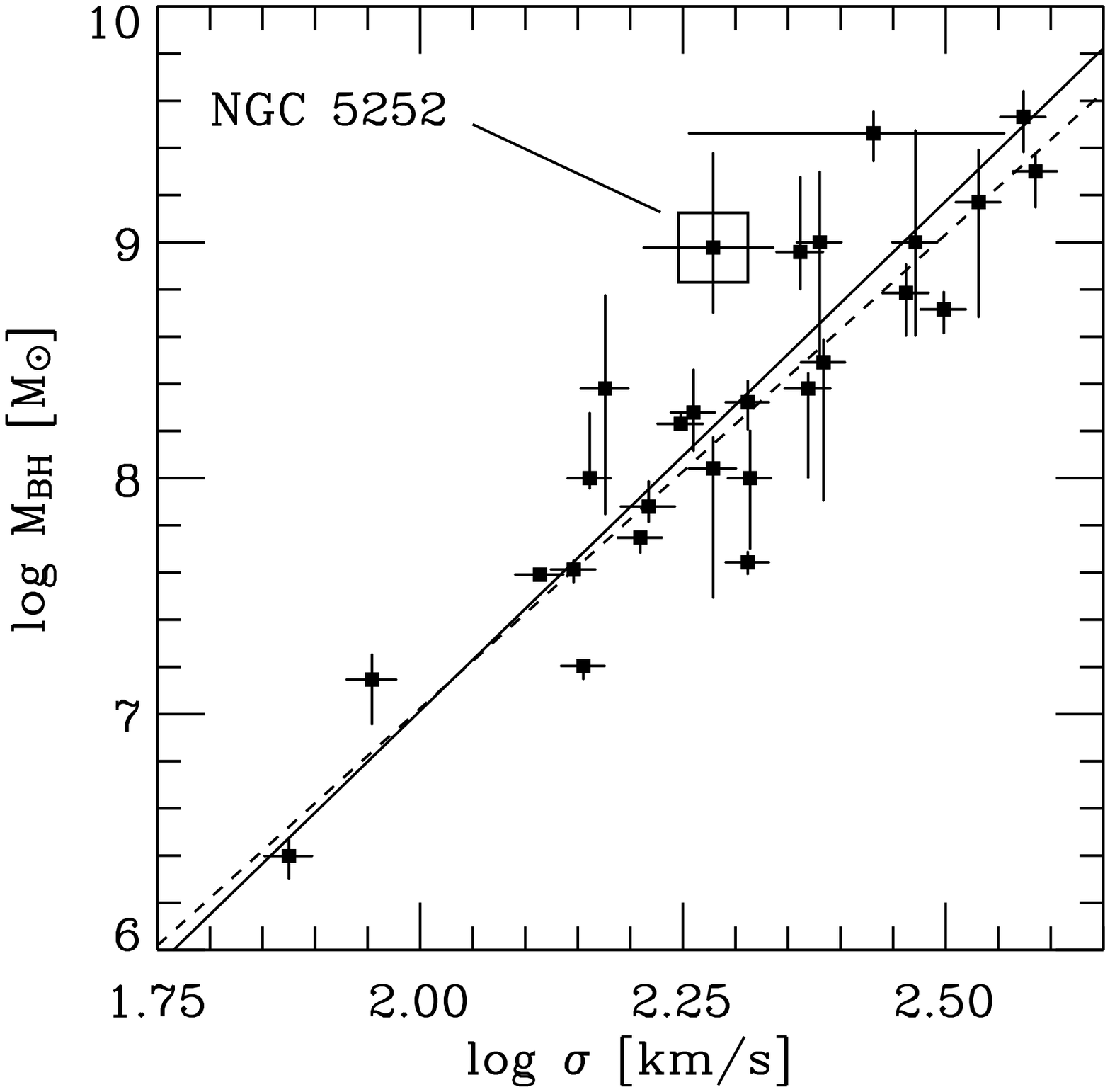,width=0.33\linewidth}
\psfig{figure=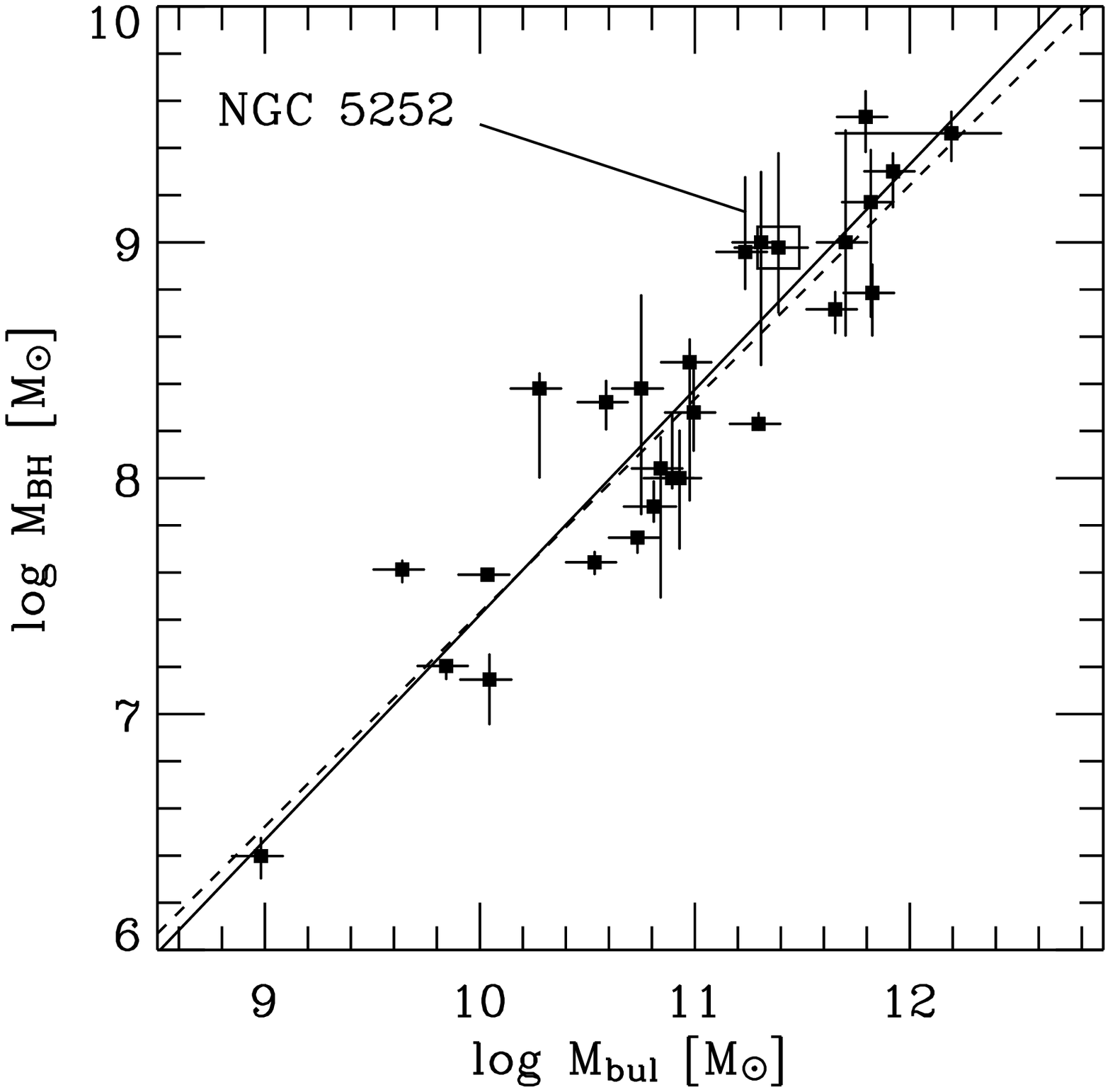,width=0.33\linewidth}
\psfig{figure=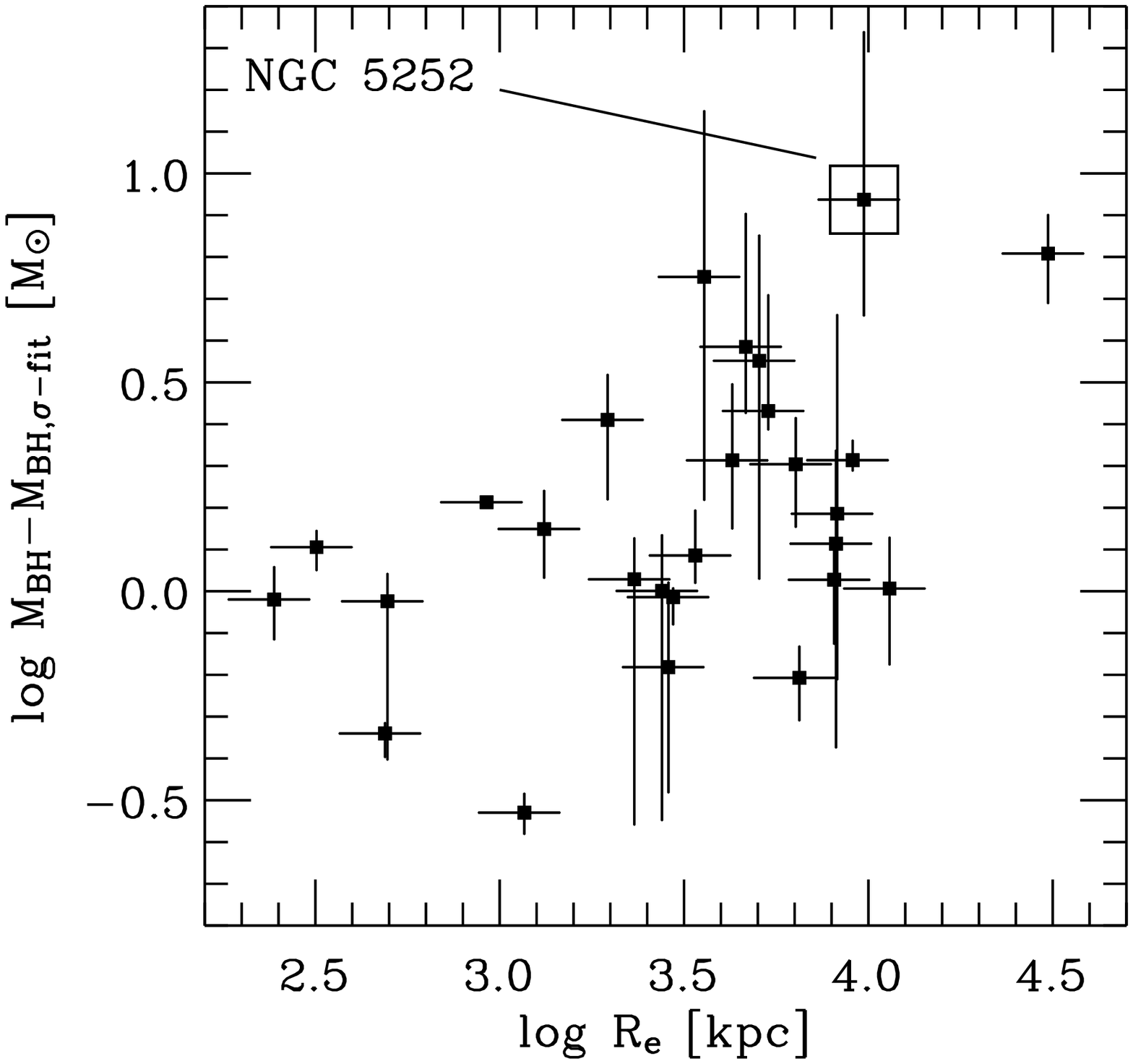,width=0.33\linewidth}}
\caption{\label{corr} 
left) M$_{BH}$ vs. stellar velocity dispersion $\sigma_c$
and middle) M$_{BH}$ vs. bulge mass with the best fits obtained from a 
bisector linear regression analysis (solid line) and ordinary least-square
(dashed line); 
right) residual of the M$_{BH}$ - $\sigma_c$ correlation vs. R$_e$ from Marconi \& Hunt (2003)}
\vskip 0.5cm
\end{figure*}

Conversely,  adopting  the central  velocity  dispersion  of NGC  5252
reported  by  Nelson \&  Whittle  (1995) of  190  $\pm$  27 \kms,  the
correlation between velocity dispersion  and black hole mass (in their
different forms  as presented by Ferrarese \&  Merritt, 2000; Gebhardt
et al.  2000; Tremaine et al.   2002; Marconi \& Hunt 2003) predicts a
mass of $M =  1.0_{-0.5}^{+1.0}\times10^8$ M$_{\sun}$, where the error
is dominated by the uncertainty in $\sigma_c$.
Taken at face value, our estimate  for the black hole mass of NGC 5252
is larger by  a factor $\sim$ 7  - 10 (depending on which  best fit is
used)   than   the  value   expected   from   this  correlation   (see
Fig. \ref{corr}). However,  combining the relatively large uncertainty
on $\sigma_c$ with  the error on $M_{BH}$ the  allowed range of upward
scatter for NGC 5252 is 0.3 - 1.2 dex, i.e. 1 - 4 times the dispersion
of the M$_{BH}$ - $\sigma_c$ correlation.

\begin{figure*}
\centerline{
\psfig{figure=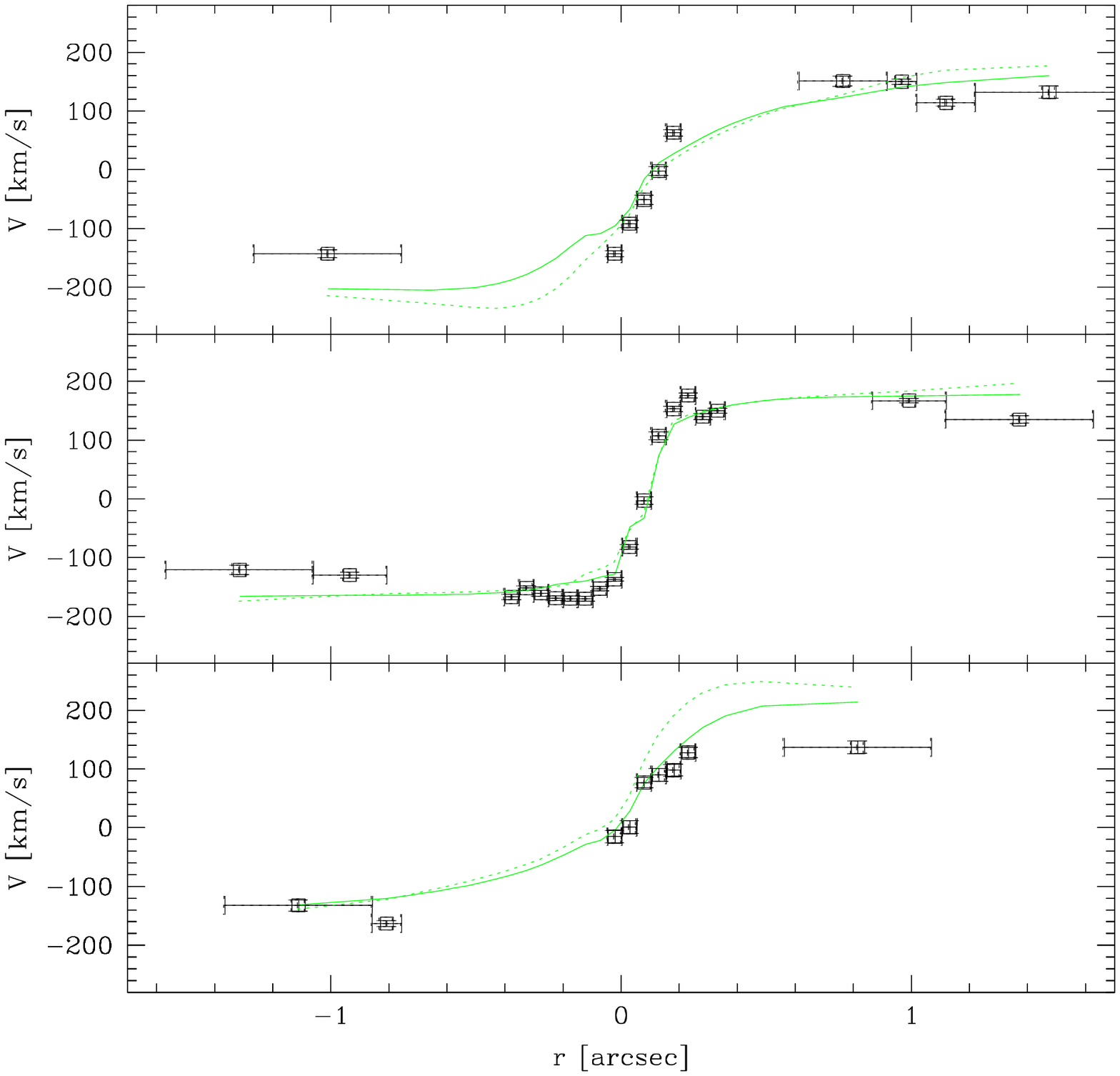,width=0.5\linewidth,angle=0}
\psfig{figure=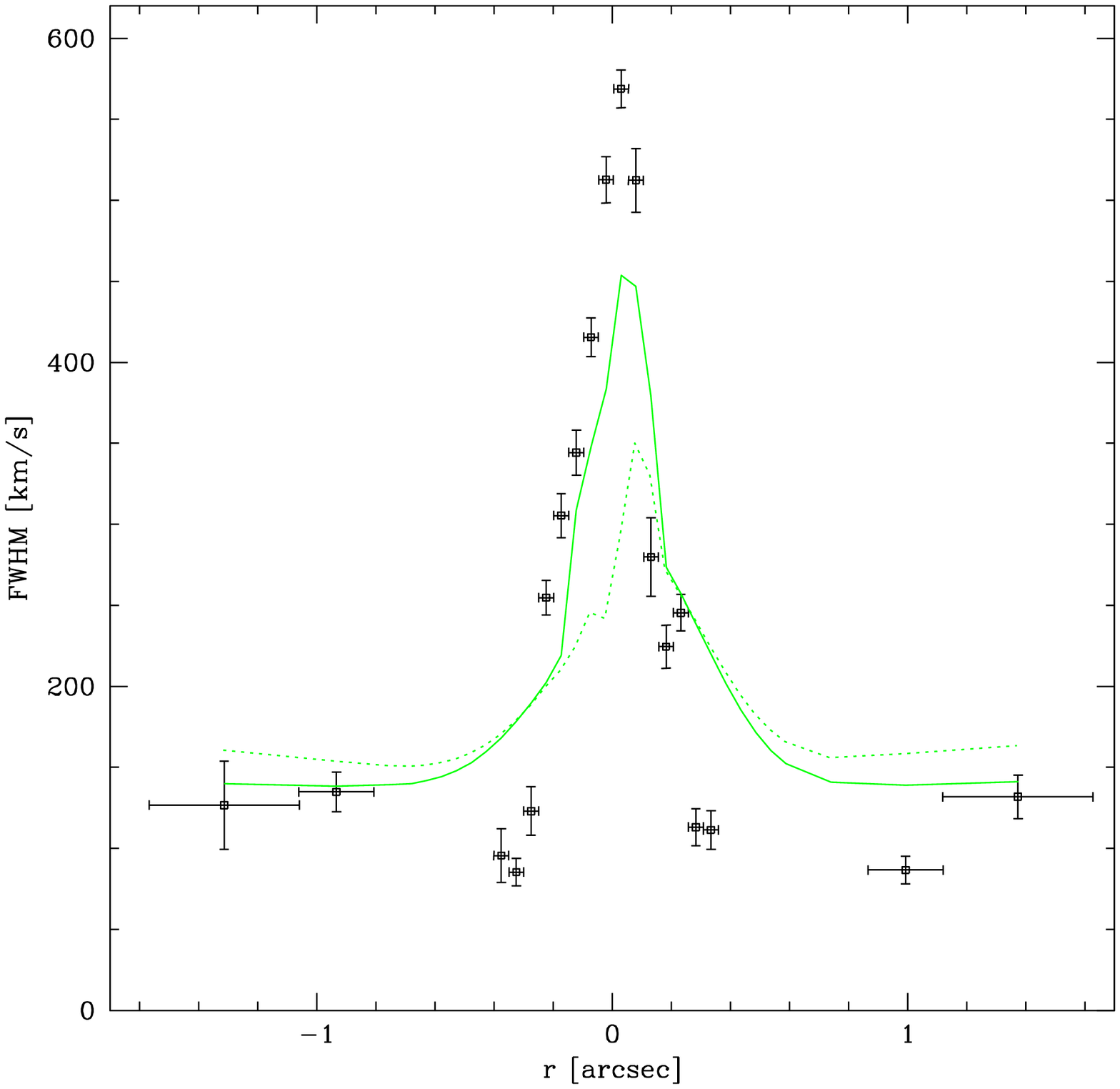,width=0.5\linewidth,angle=0}}
\vskip 0.5cm
\caption{\label{bestm8} Best fit of the rotation curves (left),
to the line surface brightness (upper right panel) 
and line width (lower right panel) obtained for 
a black mass of $M = 10^8$ M$_{\sun}$ predicted from the 
correlation between BHM and velocity dispersion.
The dotted lines correspond to the case of no central black hole.}
\end{figure*}

\begin{figure}
\centerline{
\psfig{figure=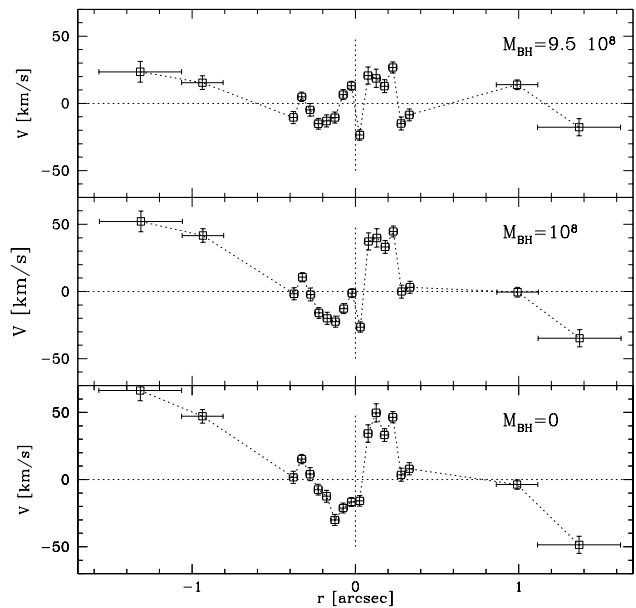,width=1.0\linewidth}}
\vskip 0.5cm
\caption{Residuals between the kinematical models and the observed velocity
points on the nuclear slit for the best fit, corresponding to 
$M_{BH}=9.5 ~10^8  M_{\sun}$ (upper panel), for a black hole mass of 
$M_{BH}= 10^8 M_{\sun}$ discussed in section 5 (middle panel) and 
for the case without central black hole (lower panel.}
\label{res}
\end{figure}

Nonetheless,  this suggests a  more critical  analysis of  our result.
The most convincing evidence in  favour of the reality of the observed
upward scatter comes from the  best fit to the rotation curves derived
fixing the BH mass to the  value predicted by the correlation (i.e. $M
=  10^8$ M$_{\sun}$),  shown  in Fig.   \ref{bestm8}.   This model  is
unsatisfactory for  several reasons:  the rotation curve  is monotonic
(it does not  show the observed fall off towards  large radii) and, in
order to fit the large nuclear gradient, it overpredicts the amplitude
of the  large scale  velocity field; secondly,  it does not  produce a
sufficiently large  line width in  the central regions.   Both effects
are clearly  indicative of the fact  that, adopting this  BH mass, the
resulting nuclear  potential well is not sufficiently  deep to account
for the observed  gas motions.  The residuals shown  in Fig. \ref{res}
(central panel) confirm the shortcoming  of this model as they present
a  relatively  large  amplitude  (up   to  50  \kms)  and,  even  more
importantly, systematic  departures on both sides of  the nucleus. The
reduced  $\chi^2$ for this  model is  1.46.  We  conclude that  a mass
larger than  predicted by the correlation  between velocity dispersion
and  black  hole mass  is  needed  in the  case  of  NGC  5252.  As  a
comparison, in the  same Fig. \ref{bestm8}, we also  show the best fit
obtained without  a central  black hole. The  drawbacks of  this model
(for which $\chi^2_r$ = 2.83) are  similar to those discussed for $M =
~10^8$ M$_{\sun}$ but quantitatively more prominent.

There are several possibilities to account for the upward scatter
in  M$_{BH}$  -$\sigma_c$  plane.  For  example,  the  measurement  of
$\sigma_c$ can be affected by  the presence of the AGN emission lines.
The  most likely  contaminants are  the [Fe  VI]$\lambda$5158  and [Fe
VII]$\lambda$5177 lines. As they lie on the wings of the Mg absorption
lines,  this  would  cause  a  decrease in  the  measured  $\sigma_c$.
Similarly,  correction  for  a  possible  rotational  contribution  to
$\sigma_c$, not  unexpected since  NGC 5252 is a S0 galaxy,  would also
enhance  the observed scatter.   An alternative  interpretation comes
from inspection of  Fig. 1 of Marconi \& Hunt  (here reproduced as the
right panel of Fig. \ref{corr}): this figure shows that the residuals
of  the M$_{BH}$  - $\sigma_c$  relation correlate  with  the galaxy's
effective radius  $R_e$.  The  brightness profile of NGC  5252 is
well described with a de Vaucouleurs  law (its Sersic index is 3.92 in
the J  band) with $R_e$ = 9.7  kpc (Marconi \& Hunt  2003), among the
largest  values for  galaxies with  measured M$_{BH}$.   In this  sense the
scatter from the M$_{BH}$  - $\sigma_c$ correlation is not unexpected,
strengthening  the  possible  role  of this  structural  parameter  in
driving the correlations between bulge and black hole properties.

\begin{figure}
\centerline{
\psfig{figure=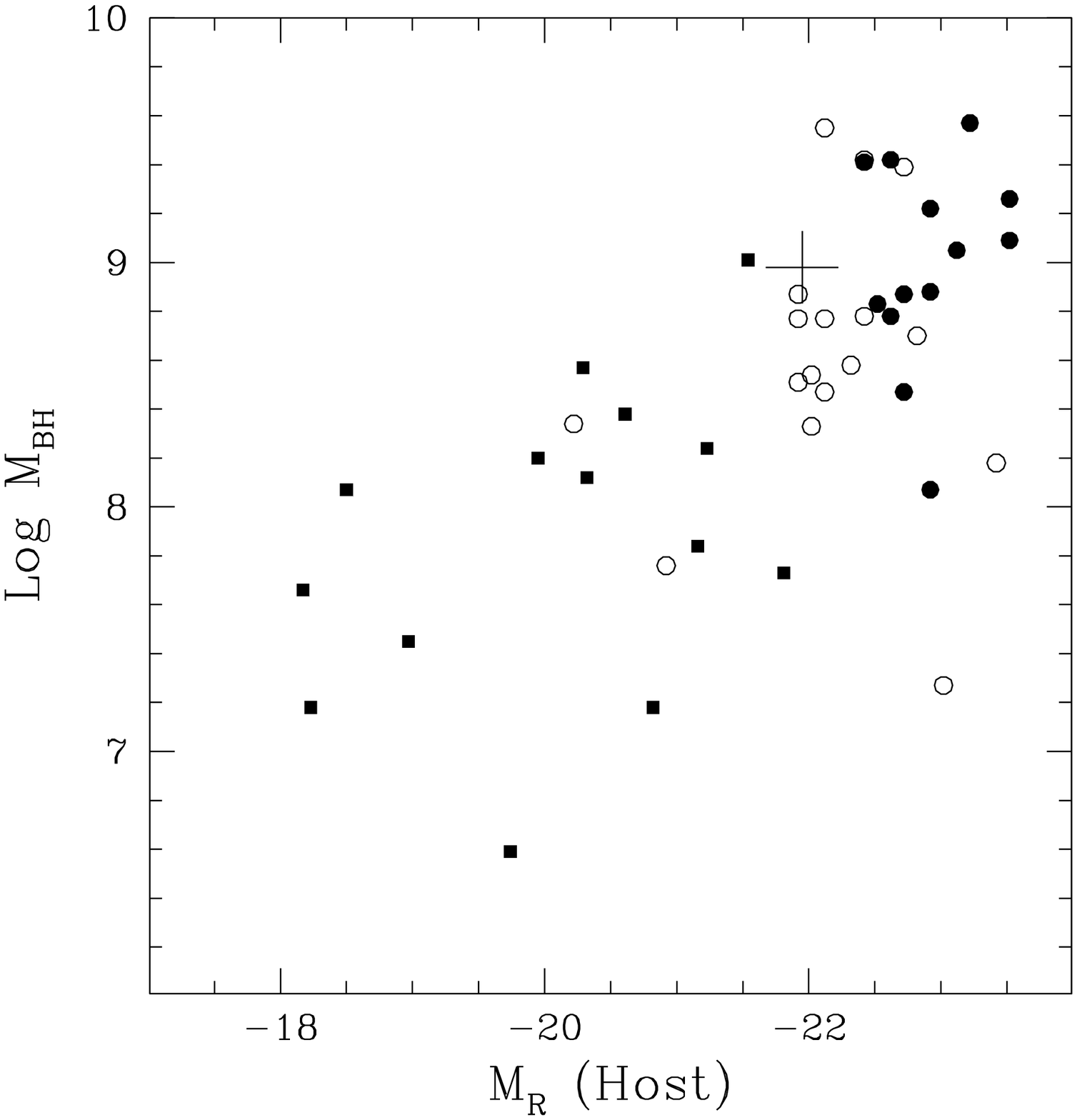,width=1.0\linewidth}}
\vskip 0.5cm
\caption{Host galaxy's magnitude versus black hole mass 
for radio-quiet QSO (open circles), radio-loud QSO (filled circles) 
and Seyfert 1 (squares) taken from McLure \& Dunlop (2001). 
The cross marks the location of NGC 5252 in this plane that falls in 
the region populated by QSO.}
\label{mclure}
\end{figure}

It is also important to compare  the properties of NGC 5252 with those
of  other active  galaxies, in  particular with  Seyfert  galaxies and
quasars  for  which estimates  of  black  hole  masses are  available,
derived combining the  widths of the broad lines  with measurements of
the BLR radius, R$_{BLR}$; BLR radii are derived in part directly from
the reverberation  mapping technique and in part  from the correlation
between  R$_{BLR}$  and  the  optical  nuclear  luminosity  (Kaspi  et
al.  2000, McLure  \&  Dunlop 2001).   At  least for  a sub-sample  of
Seyfert  galaxies it  has been  possible to  show that  adopting these
estimate they follow both the  $\sigma_c$ - M$_{BH}$ and $M_{bulge}$ -
M$_{BH}$  correlations defined  by  dynamical estimates  of BH  masses
(e.g. Wandel 2002, Peterson  2003) supporting the reliability of these
estimates.  In Fig.  \ref{mclure} we present the data  for Seyfert and
QSO  discussed   by  McLure  and  Dunlop  (2001)   including  the  new
determination of BH mass for NGC 5252.  Clearly NGC 5252 is an outlier
with respect to  Seyfert galaxies, as it harbours  a black hole larger
than  typical  for  these  objects,   but  its  host  galaxy  is  also
substantially brighter than average. On the other hand, both the black
hole  and the  bulge's mass  are typical  of the  range  estimated for
radio-quiet quasars.

The estimate of  the black hole mass in NGC 5252  allows us to explore
the nature of accretion at work in this source.  The luminosity of NGC
5252 in the hard  X-ray band 2 - 10 keV is  $\sim 2\times 10^{43}$ erg
s$^{-1}$ (Cappi  et al.  1996).  For a  standard quasars  spectrum the
luminosity  in  this band  is  $\sim$  1/30  of the  total  bolometric
luminosity (Elvis et al. 1994). As the Eddington luminosity for a 0.95
10$^9$  M$_{\sun}$  black  hole  is  L$_{Edd}$  =  1.3  $10^{47}$  erg
s$^{-1}$, we estimate  that NGC 5252 is emitting  at a fraction $\sim$
0.005 of L$_{Edd}$.   Note that, while on the one  hand, this value is
substantially  lower than  expected for  a  standard QSO  (0.1 -  0.5,
e.g. McLure \&  Dunlop 2001), on the other  it is significantly higher
than what  is derived for the low  luminosity radio-galaxies discussed
in the  Introduction.  A possible difference in  the accretion process
in NGC 5252  with respect to other Seyfert  galaxies is also supported
by its spectrum: contrarily  to the extended emission line region
where the optical line ratios are typical of Seyfert galaxies, in its
central component they  are typical of LINERs (Goncalves  et al. 1998)
and its X-ray spectrum is unusually flat ($\Gamma \sim 1.45$, Cappi et
al. 1996).

We  conclude that  NGC 5252  appears  to be  the relic  of a  luminous
quasar,  harbouring a  large mass  black  hole, that  is now  probably
accreting  at a  low  rate, rather  than  a low  black  hole mass  QSO
counterpart.

\section{Summary and conclusions}
\label{summary}

We  presented  results from  HST/STIS  long-slit  spectroscopy of  the
Seyfert 2 galaxy  NGC 5252 at a redshift z =  0.023. The nuclear
velocity  field  shows  a   general  reflection  symmetry  and  it  is
consistent with the presence of gas in regular rotation; however there
are superposed patches of  disturbed gas, localized on two off-nuclear
emission-line blobs. We separated these two components on the basis of
the behaviour of line width and flux, discarding the regions where the
disturbed gas appears to dominate. The dynamics of the rotating gas
can then  be accurately reproduced  by motions in  a thin disk  when a
compact  dark  mass  of  $M_{BH}  =  0.95_{-0.45}^{+1.45}\times  10^9$
M$_{\sun}$, very  likely a  supermassive black hole,  is added  to the
stellar mass component; the  estimate of $M_{BH}$ is quite stable
against different choices of the spatial regions modeled.  This result
provides  {\sl  a posteriori}  support  for  our  interpretation of  a
gravitational origin for the observed gas motion, also considering the
internal consistency  of our  modeling, e.g. the  value of  the galaxy
mass-to-light ratio, the reduced  velocity gradient in the off-nuclear
slits with respect  to what is measured on  the nucleus.  Nonetheless,
we cannot rule  out alternative explanations only on  the basis of the
data presented in this paper.

The M$_{BH}$ value thus estimated is in good agreement with the correlation
between bulge  and BH mass, while  it is larger that  predicted by the
correlation proposed  between BH mass vs  velocity dispersion ($M_{BH}
\sim 10^8$), with an upward scatter  between 0.3 and 1.2 dex, i.e. 1 -
4  times the  dispersion of  the correlation.   We show  that adopting
$M_{BH} \sim  10^8$ it is impossible  to obtain a  satisfactory fit to
the data.   A possible interpretation  of this scatter comes  from the
presence of a correlation between  the {\sl residuals} of the M$_{BH}$
- $\sigma_c$  relation with  the galaxy's  effective radius  $R_e$. In
fact NGC 5252 has one of the largest values of $R_e$ for galaxies with
measured  M$_{BH}$, suggesting  a possible  role of  this parameter  in the
connection between bulge and black hole properties.

Concerning  the properties  of  its  active nucleus,  NGC  5252 is  an
outlier when compared to the  available data for Seyfert galaxies, not
only  as  it harbours  a  black hole  larger  than  typical for  these
objects, but  also as its  host galaxy is substantially  brighter than
average for Seyfert galaxies.  On  the other hand, both the black hole
and the bulge's mass are typical of the range for radio-quiet quasars.
Combining the  determined BH mass  with the hard X-ray  luminosity, we
estimate  that NGC  5252 is  emitting at  a fraction  $\sim$  0.005 of
L$_{Edd}$. This active nucleus thus  appears to be a quasar relic, now
probably accreting at  a relatively low rate, rather  than a low black
hole mass counterpart of QSOs.

\end{document}